\documentclass[reprint,amsmath,amssymb,aps,pra,superscriptaddress,nofootinbib,twocolumn,floatfix,nolongbibliography]{revtex4-2}
\usepackage{microtype}
\usepackage{comment}

\usepackage[USenglish]{babel}
\usepackage[T1]{fontenc}
\usepackage{hyperref}
\usepackage{graphicx}
\usepackage{array}[=2016-10-06]

\usepackage{bbm}
\usepackage{mathtools}
\usepackage{amsfonts}
\usepackage{mathrsfs}
\usepackage{physics}
\usepackage{slashed}
\usepackage{tensor}
\usepackage{amsmath}
\usepackage{amssymb}
\usepackage{multirow}
\usepackage{amsfonts,dsfont}
\usepackage{bm}
\usepackage{physics}
\usepackage{mathtools}
\usepackage{slashed}

\usepackage{graphicx}
\usepackage[dvipsnames]{xcolor}
\usepackage{rotating}
\usepackage[caption=false,singlelinecheck=false]{subfig}
\usepackage{float}
\usepackage{placeins}
\usepackage{booktabs}
\usepackage{makecell}
\usepackage{tabstackengine}

\usepackage{dcolumn}
\usepackage{bm}
\usepackage{braket}
\usepackage{xcolor}
\usepackage{bbold}
\usepackage{dsfont}
\usepackage{mathrsfs}
\usepackage{float}

\usepackage{xspace}
\usepackage{siunitx}
\usepackage{xfrac}
\usepackage{hyperref}
\usepackage[nameinlink]{cleveref}
\usepackage{appendix}

\usepackage{xifthen}
\usepackage{xcolor}
\usepackage{enumitem}
\hypersetup{
	colorlinks,
	linkcolor={red!75!black},
	citecolor={red!75!black},
	urlcolor={blue!75!black}
}

\usepackage[utf8]{inputenc}

\graphicspath{{figures/}}

\definecolor{revtexred}{HTML}{bf0000}

\begin{document}

\preprint{APS/123-QED}

\title{Generative sampling with physics-informed kernels}
\author{Friederike Ihssen}
\author{Renzo Kapust}
\email{kapust@thphys.uni-heidelberg.de}
\affiliation{
    Institute for Theoretical Physics, Universit\"{a}t Heidelberg,
Philosophenweg 16, D-69120, Germany
}

\author{Jan M. Pawlowski}
\affiliation{
    Institute for Theoretical Physics, Universit\"{a}t Heidelberg,
Philosophenweg 16, D-69120, Germany
}%
\affiliation{
    ExtreMe Matter Institute EMMI, GSI, Planckstr. 1, D-64291 Darmstadt, Germany
}


\begin{abstract}
	
We construct a generative network for Monte-Carlo sampling in lattice field theories and beyond, for which the learning of layerwise propagation is done and optimised independently on each layer. 
The architecture uses physics-informed renormalisation group flows that provide access to the layerwise propagation step from one layer to the next in terms of a simple first order partial differential equation for the respective renormalisation group kernel through a given layer. 
Thus, it transforms the generative task into that of solving once the set of independent and linear differential equations for the kernels of the transformation. As these equations are analytically known, the kernels can be refined iteratively. This allows us to structurally tackle out-of-domain problems generally encountered in generative models and opens the path to further optimisation. We illustrate the practical feasibility of the architecture within simulations in scalar field theories. 

\end{abstract}

\maketitle

\section{Introduction}
\label{sec:Introduction}

Generative networks for Monte-Carlo simulations on high-dimensional data spaces with a given distribution face two out-of-domain (OOD) problems that hamper progress in this area. 

The most obvious one is related to the high dimensionality of the data space that typically renders the learned distributions inaccurate. Roughly speaking this originates in the fact that interpolation problems in high dimensions are equally ill-conditioned as extrapolation problems in low dimensions. Consequently the computational costs of the final sampling steps with the true distribution scale badly with the dimension, which is a manifestation of the curse of dimensionality. 

The second problem is a standard extrapolation problem and may be explained as follows: assume that the network has been trained up to a given sample size related to a limited information content: the size of a sample is in one-to-one correspondence to the access to a specific set of moments or cumulants of the distribution with the orders $n<N_\textrm{max}$, where $N_\textrm{max}$ increases with the sample size but scales very badly. For instance, for a scalar lattice field theory in two space-time dimensions, the dimensionality of the data space is given by $V = N_\tau\times N_s$, where $N_\tau$ is the number of lattice points in the (Euclidean) time direction and $N_s$ is the number of lattice points in the spatial direction. In large scale lattice simulations, the lattice volume $V$ easily reaches $10^6$ to $10^8$, depending on the space-time dimension. 

Further sampling/training gives access to higher order moments/cumulants with $n>N_\textrm{max}$. However, the information carried by the higher order moments, or rather their irreducible parts, is not contained in the lower ones. This information is provided by the final accept-reject step with the true distribution which hence is increasingly expensive: it scales as simply sampling with the true distribution in the first place. These two OOD problems lead to the bad scaling properties in the limit of large dimensions observed in many generative networks.

We hasten to add that the above problems are absent for distributions which have a finite number of independent moments after a reformulation within appropriate degrees of freedom but have a seemingly infinite number of independent moments in the given degrees of freedoms. If a generative network can unravel such a typically highly non-linear transformation within a rather finite data or training set, the generative task is practically solved. 

In turn, the general situation asks for the construction of networks or sampling algorithms for which the problems above are solved constructively rather than hoping for the above underlying simplicity. Such a framework has to use the structural information that is encoded in the normalised statistical true distribution $p(\varphi)$, 
\begin{align} 
	p(\varphi) = \frac{1}{{\cal N}} \,e^{-\hat S(\varphi)}\to  e^{-S(\varphi)}\,,
\label{eq:GenDistr}
\end{align} 
with the normalisation ${\cal N}$ and the exponential measure factor $\hat S(\varphi)$. In the second step in \labelcref{eq:GenDistr} we have absorbed the normalisation in a shift of the action, $S= \hat S +\log \mathcal{N}$. Then, $S$ is nothing but the negative log likelihood and contains all the structural  information of the distribution. In physics applications $\hat S(\varphi)$ is also called the action or effective Hamiltonian and $\varphi$ is a field. For general applications, $\varphi$ is a map from the set of data points ${\cal D}$ to the set of data values ${\cal T}$ and both sets may be discrete or continuous. 

In the present work we suggest such a generative architecture based on physics-informed renormalisation group flows \cite{Ihssen:2024ihp} (PIRGs) which aims at solving constructively the two OOD problems discussed above. PIRGs offer a comprehensive and general framework for invertible transformations of high-dimensional distributions with the RG-time $t$, analogously to the layerwise propagation and transformation in invertible networks. Importantly, all these steps follow tractable analytic differential equations which is key for the resolution of the OOD problems. 

\begin{figure}[t]
    \centering
    \includegraphics[width=0.5\textwidth]{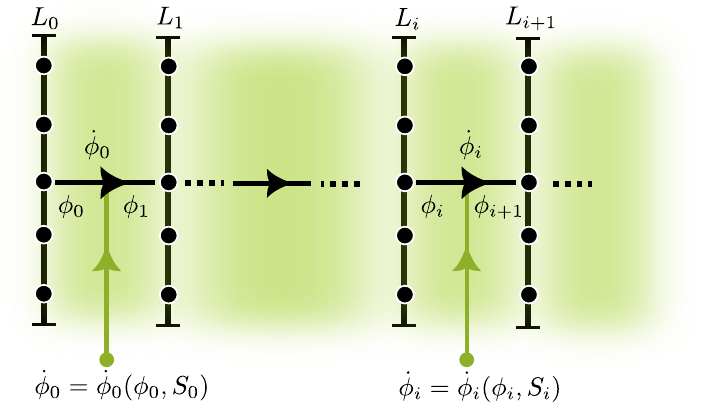}
    \caption{Illustration of the construction principle of PIKs for degrees of freedom $\phi$ that are propagated through the layers $L_i$ using independent kernels $\dot{\phi}_i$. \hspace*{\fill}}
    \label{fig:ConceptionalIntroduction}
\end{figure}

We proceed with a rough explanation of the construction principle with the example of a lattice field theory in mind. For this purpose we consider a deep network of layers $L_i$ with variables $\phi_i$ on the layer or rather the respective lattice. This is depicted in \Cref{fig:ConceptionalIntroduction}.

The PIRG approach allows the construction of maps $\phi_i\to \phi_{i+1}$ on the basis of the pair $\{S_i(\phi_i), \phi_i\}$ of the negative log likelihood $S_i$ and the field $\phi_i$ on a given layer $L_i$. The input layer $L_0$ has a chosen pair $\{S_0(\phi_0), \phi_0\}$. Natural choices for the input pair are the full action or negative log likelihood of the true distribution with  $\phi_0=\varphi$ and $S_0=S$ or a Gau\ss ian distribution with $S_0=S_\textrm{free}$, where the subscript indicates a quadratic action. The kernels of these transformations are \textit{physics-informed kernels} (PIKs). The respective transformation of the field $\phi_{i}$ on the layer $L_i$ to the field $\phi_{i+1}$ on the layer $L_{i+1}$ is given by a renormalisation group equation for the pair $\{S_i(\phi_i), \phi_i\}$, which is a diffusion equation for $S_i$: for the negative log likelihood $S_i$ this is the Wegner equation \cite{Wegner_1974}, for other generating functions similar analytic RG equations exist, e.g. the generalised RG equation \cite{Pawlowski:2005xe} for the rate function. 

The key point of the PIRG approach is its use of the diffusive RG equation as a differential equation for the pair $\{S_i(\phi_i), \phi_i\}$ instead of a diffusive equation for $S_i$. For the present generative architecture this freedom is used as follows: Rather than following the diffusive dynamics and evolving the action $S_i$, all $S_i$'s are fixed analytically. Then, the underlying renormalisation group equation is a simple linear differential equation for the kernel. This gives us an analytic access to the map between different layers in a deep network. Importantly, this linear differential equation does not contain information from the previous layers, thus completely decoupling the layerwise propagation steps from each other. With this physics-informed decoupling structure and determination of the layerwise map by an RG equation, the sampling network is to our knowledge qualitatively different from any other networks. The possible trivial parallelisation as well as the nature of the analytically known diffusion equation on the layers structurally resolve both of the OOD problems mentioned in the beginning. Furthermore, the underlying RG structure also allows for a systematic and optimisable upgrade of this procedure by also allowing for a learning of the set of actions $\{S_i\}$ and hence the construction of optimised PIKs, also accommodating improved or perfect actions in the RG procedure~\cite{Holland:2025fsa}. 

In the present work we briefly review PIRGs in the light of the present application to generative architectures in \Cref{sec:PIRGsNutshell}. The  general PIK architecture is put forward in \Cref{sec:PIKs}, where it is argued that PIKs allow for the construction of `truly' generative networks. Here `truly' refers to the fact that the probability of the final accept-reject step stays close to unity. The conceptual developments are showcased within a zero-dimensional lattice field theory as well as within first steps in higher dimensions in \Cref{sec:PIKsAtWork}. We conclude in \Cref{sec:Outlook} with a summary and an outlook of some of the natural applications of PIKs. Some technical details are deferred to the Appendices, where also some of the machine-learning driven optimisation aspects are discussed.

\section{Physics-informed renormalisation group flows in a nutshell}
\label{sec:PIRGsNutshell}

Renormalisation group flows describe the scale-dependence and general reparametrisations of a given statistical distribution. While developed in quantum and statistical field theories, the concept is far more general. Both aspects are governed by the infinitesimal change of the distribution with the renormalisation time $t$. Accordingly, instead of the field $\varphi$ in \labelcref{eq:GenDistr}, we consider fields $\phi_t$ that are related to $\varphi$ with a general non-linear transformation. Importantly, we do not have to resolve this transformation, but only its infinitesimal change from $\phi_t \to \phi_{t\pm\epsilon}(\phi_t)$. This transformation comes with a change of the distribution $p_t(\phi)$ and the respective differential equation is either derived for the measure $p_t(\phi)$ or its Laplace or Fourier transform 
\begin{subequations} 
	\label{eq:Zpphi}
\begin{align} 
Z(J)= \int {\cal D} \phi \,p_t(\phi) \,e^{ \sum\limits_{\hat n} \phi_{\hat n} J_{\hat n}}  \,, \quad {\cal D}\phi=\prod_{\hat n\in {\cal D}}\int_{{\cal T}_t} d\phi\,, 
\label{eq:ZpphiEq}
\end{align}
where 
\begin{align} 
	p_t(\phi) = e^{-S_t(\phi)}\,. 
	\label{eq:p(phi)}
\end{align}
\end{subequations} 
In \labelcref{eq:Zpphi} we have introduced the negative log likelihood $S_t(\phi)$, and the fields $\phi$ take values in ${\cal T}_t$. In general, $\hat n\in{\cal D}$ labels data points in the discrete or continuous data space. We have restricted ourselves to the case of discrete data points in \labelcref{eq:Zpphi}, such as also present in our lattice field theory examples. There, data points are sites on a rectangular lattice with $\hat n =(n_1,n_2,...,n_d)$ and $n_i\in [0,1,...,N_{i,\textrm{max}}]$. The field $\phi_{\hat n}$ takes real values on each of these lattice sites. In the present work we use a (hyper-) cubic lattice with $N_{i,\textrm{max}}=N_\textrm{max}$ for all $i=1,...,d$. Then, the field is a map 
\begin{align} 
\phi: {\cal D}\to \mathbbm{R}^{\cal D}\,,\qquad 	{\cal D} = [0,1,...,N_{\textrm{max}}]^d\,, 
\label{eq:phiMap}
\end{align} 
with $\phi \in {\cal C}$, and ${\cal C}$ is the space of maps \labelcref{eq:phiMap} with a given norm, commonly the $L_2$-norm. Typical large lattice sizes range between $N_\textrm{max} \propto 10^3$ in d=1,2,3 and $N_\textrm{max}\propto 10^2$ for $d=4$. The most general scale and reparametrisation RG transformation is accommodated by the flow of the measure 
\begin{align} 
	\frac{d \,p_t(\phi)}{d t} = \frac{\partial}{\partial\phi }\Bigl[ \Psi_t(\phi) \, p_t(\phi)\Bigr]\,, \quad p_t(\phi) = e^{-S_t(\phi)} \,, 
\label{eq:Wegner74} 
\end{align} 
and was put forward in \cite{Wegner_1974}. Recently, the Wegner flow has received attention within applications in generative models and its relation to optimal transport~\cite{Cotler:2022fze, Mate:2023fxa, Cotler:2023lem, Albergo:2025nets, Sheshmani:2025ylv}. The flow in \labelcref{eq:Wegner74} is a total derivative and hence leaves the total distribution $Z(0)=1$ invariant. In the absence of boundary terms and for $d {\cal T}_t/dt=0$ this invariance is given by 
\begin{align} 
\frac{d \,Z(0)}{d t}   \equiv 0\,. 
\label{eq:dtZ0}
\end{align} 
The case $d {\cal T}_t/dt \neq 0$ has been considered in \cite{Ihssen:2024ihp} in the rate function formulation of PIRGs and is readily mapped to the case $d {\cal T}_t/dt=0$. 

An important subgroup of transformations are the choices $\Psi_t(\phi) \propto  \partial S_t(\phi)/\partial\phi$ with the 
RG-time dependent classical action $S_t$. Then, \labelcref{eq:Wegner74} takes the form of a Fokker-Planck equation in high-dimensions, and is related to the Langevin evolution of the field (Stochastic Quantisation) and the Polchinski equation \cite{Polchinski1984} for the Wilsonian effective action. \Cref{eq:Wegner74} with or without the source term carries redundancies which may slow down and complicate its numerical solution. For this and further structural, stability, and numerical convenience reasons one typically considers the rate function or effective action $\Gamma(\phi)$ obtained from a Legendre transform of $\ln Z(J)$. The respective generalised RG equation reads \cite{Pawlowski:2005xe}, 
 \begin{align} 
 \left(	\frac{d}{d t}  +\dot\phi_t \frac{\partial}{\partial\phi} \right)\Gamma_{\hspace{-.07cm}t}= \frac12 \Tr  \left[\frac{1}{\frac{\partial^2\Gamma_{\hspace{-.07cm}t}}{\partial\phi^2}+R_t} \left(\frac{d}{d t} +  2 \frac{\partial \dot\phi_t}{\partial\phi}\right)R_t \right], 
 \label{eq:GenRG}
 \end{align}
with the $\phi$-independent scaling-kernel $R_t$ and the field transformation $\dot\phi_t(\phi)$. For $\dot\phi_t=0$ it reduces to the Wetterich equation \cite{Wetterich:1992yh}, see also \cite{Ellwanger:1993mw, Morris:1993qb}. If we set the change of the scaling kernel to zero, $d R_t/dt =0$, the two equations \labelcref{eq:Wegner74,eq:GenRG} are related by the choice $\Psi\propto \dot\phi_t$. For more details see \cite{Ihssen:2022xjv, Ihssen:2023nqd, Ihssen:2024ihp}. Note also that in the functional RG literature the total RG-time derivative $d/dt$ is commonly written as $\partial_t$.  

The key idea underlying physics-informed RG flows \cite{Ihssen:2024ihp} is to exploit the full generality of the Wegner flow \labelcref{eq:Wegner74} for the distribution, the generalised flow \labelcref{eq:GenRG} for the rate function, or related ones for other generating functions: Instead of solving the flows for a given $p_t(\phi), \, \Gamma_{\hspace{-.07cm}t}(\phi)$ or other generating function $F_t$, we keep the coordinates $\phi$ general and solve the equations for pairs 
\begin{align}
	\bigl(F_t(\phi)\,,\,\dot\phi_t(\phi)\bigr)\,,\qquad \textrm{with}\qquad F_t\in\{S_t,\Gamma_{\hspace{-.07cm}t},...\}\,. 
	\label{eq:PIRG-Pair}
\end{align}
Here, $\dot\phi_t(\phi)$ is the local change at a given RG-time $t$, and in the examples considered $F_t$ is either the negative log likelihood or the rate function. For the present purpose we take the RG-time in a normalised time interval,  
\begin{align}
t \in I\qquad \textrm{with}\qquad 	I=[0,1]\,.
	\label{eq:INormalised}
\end{align}
Here, $t=0$ is the initial time and $t=1$ is the final time. 
In particular, \labelcref{eq:PIRG-Pair} with \labelcref{eq:Wegner74,eq:GenRG} also accommodates solving \labelcref{eq:Wegner74,eq:GenRG} for $\dot\phi_t$ for a given negative log likelihood or action $S_t$ or rate function $\Gamma_t$. 

An illuminating example for a flowing pair $\bigl(p_t(\phi)\,,\dot\phi_t(\phi)\bigr)$ is the pair that leads to 
normalising flows in lattice field theories~\cite{Luscher:2009eq, Rezende:2016vin, Albergo:2019eim, Nicoli:2020njz}. More specifically, we consider an inverse normalising flow with the Gau\ss ian action $S_0=S_\textrm{free}$ at $t=0$ and the full action $S_1=S$ at $t=1$. 
This constitutes a map from a free Gau\ss ian theory with the normalised distribution $p_\textrm{free}(\phi)= \exp\{-S_\textrm{free}(\phi)\}$ to the theory with full distribution $p(\varphi)$ in \labelcref{eq:GenDistr}. In terms of cumulants, normalising flows map theories with infinitely many independent cumulants to a theory with only one independent cumulant, the second order one. Evidently, in this case the map accommodates an infinite amount of information. This is directly visible for the rate function $\Gamma_{\hspace{-.07cm}t}$, where normalising flows constitute a flow from the rate function of the full theory with infinitely many non-trivial cumulants to a Gau\ss ian, quadratic rate function with only one cumulant; all higher orders are absorbed in the map $\dot\phi$ via \labelcref{eq:GenRG}. 

The PIRG setup with its analytic flows allows us to include the additional - infinite - amount of information within the sampling procedure beyond the training. The key novel feature of the architecture is a structural property of the physics-informed RG flows: If the \textit{target action} $F_t\in\{S_t, \Gamma_{\hspace{-.07cm}t},...\}$ is fully specified and known, the associated map $t \to F_t$ is completely independent of the previous ones and only depends on the target action at time $t$,
\begin{align} 
	\dot\phi_t(\phi)= \dot\phi_t(\phi;F_t)\,.
	\label{eq:PIRG-dotphi}
\end{align} 
Put differently, the \textit{kernel} $\dot\phi_t$ of the map is \textit{physics-informed}. This perspective has also been used in~\cite{Albergo:2025nets}, where \labelcref{eq:PIRG-dotphi} is solved within an architecture with a global (for all $t$) loss function. In the present work we use that \labelcref{eq:PIRG-dotphi} leaves us with independent tasks of learning or solving $\dot \phi_t$ for the whole time line from t=0 to t=1. These separate tasks represent standard RG-steps and hence carry the respective physics information. They can be trivially parallelised and also allow for an independent iterative optimisation. Importantly, the setup structurally avoids the curse of dimensionality due to the analytic access. 

We close this brief review with the remark, that in the present work we introduce and illustrate this novel simulation architecture using flows of $p_t(\phi)$ with the Wegner equation \labelcref{eq:Wegner74}. The also highly relevant discussion of the qualitative reduction of redundancies achieved by using the rate function or related generating functions as well as the further key question of full optimisation within these models is deferred to future works.

\section{Sampling architecture with physics-informed kernels}
\label{sec:PIKs}

In the present Section we use the PIRG setup to construct general flows with \textit{physics-informed kernels} (PIKs) $\dot\phi_t$, \labelcref{eq:PIRG-dotphi}, in the space of distributions. The general architecture of PIKs is put forward in \Cref{sec:GenArchitecture} at the example of the PIRG pair $(S_t,\dot\phi_t)$ and the Wegner flow \labelcref{eq:Wegner74}. Possible parametrisations of PIRG pairs and solution strategies are discussed in \Cref{sec:GenParaSt,sec:GenParadotphi}. Finally, we outline optimisation procedures in \Cref{sec:Optimisation}.

\subsection{General architecture}
\label{sec:GenArchitecture} 

With PIKs we aim at building a generative architecture that avoids the two OOD problems discussed in the introduction, which would enable the efficient sampling with sample sizes orders of magnitude beyond the sample size generated and used in the training. The architecture uses the PIRG pairs to enforce a path for the negative log likelihood or action $S_t$, leaving us with the task of computing or learning the respective map $\dot\phi_t$ from its analytically known differential equation.

In the following, we start from a simple base distribution $p(\varphi)$ at $t=0$ and flow into a more complex target distribution at $t=1$. Since the architecture is based on the flows \labelcref{eq:Wegner74,eq:GenRG}, it is invertible as the flows are invertible. For the chosen flow direction, the partition sum at $t=0$ reads 
\begin{align}
    Z_0 = \int \mathcal{D}\varphi \; p(\varphi)\,,\qquad \phi_{0} = \varphi\,, 
  \label{eq:BaseDistribution}
\end{align}
and the flow of the field $\varphi$ is given by 
\begin{align}
    \frac{d\phi_t}{dt} = \dot{\phi_t}(\phi_t)\,, \qquad \phi_{t=0} = \varphi\,. 
\label{eq:FieldODE}
\end{align}
In \labelcref{eq:FieldODE}, the physics-informed kernel vector field 
\begin{align} 
\dot{\phi}_t: I \times {\cal C}\to {\cal C}\,, 
\label{eq:dotPhi} 
\end{align} 
characterises the transformation at each time $t\in I$ with the normalised RG-time interval $I=[0\,,\,1]$. As noted before, the partition sum is invariant under the transformation \labelcref{eq:Wegner74} and hence $Z_t$ is $t$-independent, see \labelcref{eq:dtZ0}. We derive with \labelcref{eq:Wegner74} that 
\begin{align}
    Z = \int \mathcal{D}\phi \,p_t(\phi)\,, \qquad \frac{ d Z}{dt} =  \int \mathcal{D}\phi \,\frac{ d \,p_t(\phi)}{dt}=0\,. 
\label{eq:WegnerEvolutionIntegrand}
\end{align}
In \labelcref{eq:WegnerEvolutionIntegrand}, the total derivative nature of \labelcref{eq:Wegner74} has been used as well as the assumption of the absence of boundary terms. The total change of field and action or negative log likelihood is covered by 
\begin{align}
	\frac{d S_t(\phi)  }{dt} +\dot{\phi}_t(\phi) \,\frac{\partial}{\partial \phi} S_t(\phi)= \frac{\partial}{\partial \phi} \, \dot{\phi}_t(\phi) \,,
	\label{eq:WF-dotphi}
\end{align}
which is the Wegner equation for the action, readily deduced from \labelcref{eq:Wegner74}. Note that the right-hand side is nothing but the total derivative of $S_t(\phi_t)$. Here, $\partial/\partial\phi$ denotes the partial derivative w.r.t.\ each component in $\phi$.

A key observation behind PIKs is the following: The PIRG setup \labelcref{eq:PIRG-Pair,eq:PIRG-dotphi} allows us to read \labelcref{eq:WF-dotphi} independently for each $t$ as a simple linear first order differential equation for $\dot\phi_t(\phi)$ with a fully determined and accessible trajectory $S_t$. The kernel $\dot{\phi}_t$ then accommodates the change of the coordinate system in the data space that is induced by this trajectory. This structure also underlies other formulations of PIKs with $S_t\to F_t$, i.e.~based on the PIRG pair with the rate function, $(\Gamma_t,\dot\phi_t)$. Then, \labelcref{eq:WF-dotphi} for $\dot\phi_t$ is substituted by \labelcref{eq:GenRG} and interpreted as a simple first order differential equation for $\dot\phi_t(\phi;\Gamma_t)$, see the discussion around \labelcref{eq:PIRG-dotphi}.

Accordingly, to our knowledge, the PIK architecture differs qualitatively from standard generative networks that aim to improve Monte Carlo simulations. This is due to the existence of accessible analytic maps for the layerwise propagation of the PIRG pair 
\begin{align} 
	(F_{t_i}\,,\,\dot\phi_{t_i})\to  (F_{t_{i+1}}\,,\,\dot\phi_{t_{i+1}})\,,
\end{align}
with its independent solutions $\dot\phi_{i}(\phi; F_{t_i})$ with \labelcref{eq:WF-dotphi,eq:GenRG} or respective equations for general functions $F_t$.

\begin{figure}[t]
    \centering
    \includegraphics[width=0.45\textwidth]{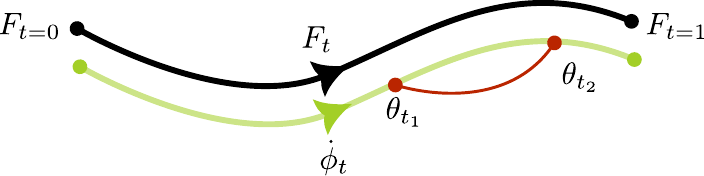}
    \caption{Training of most generative models from a distribution characterised by $F_{t=0}$ to one characterised by $F_{t=1}$. The path $F_t$ is not fully determined and accessible. Optimisable parameters $\theta_t$ at different times are entangled, which is indicates by the red entanglement line between the parameters $\theta_{t_1}$ and $\theta_{t_2}$. \hspace*{\fill}}
    \label{fig:TraditionalLearning}
\end{figure}

\begin{figure}[t]
	\centering
	\includegraphics[width=0.45\textwidth]{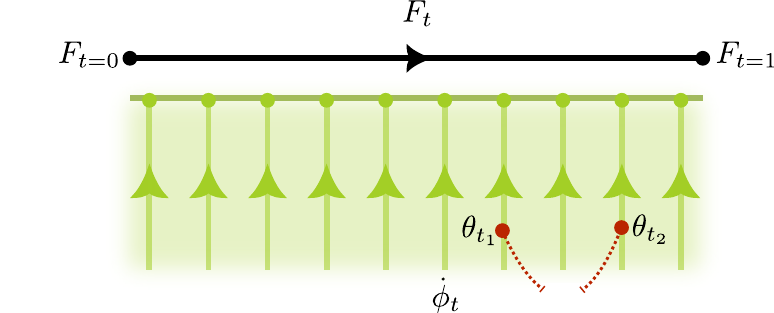}
	\caption{Training of PIKs from a distribution characterised by $F_{t=0}$ to one characterised by $F_{t=1}$. The path $F_t$ is fully determined and accessible. Optimisable parameters $\theta_t$ at different times are not entangled, which is indicated by the broken dashed red lines.  \hspace*{\fill}}
	\label{fig:PIKLearning}
\end{figure}

In the following, we contrast the learning structure of standard generative approaches, depicted in \Cref{fig:TraditionalLearning} with that of PIKs, depicted in \Cref{fig:PIKLearning}. Here the black line symbolises the path $S_t(\phi)$ connecting the base and targeted action. The green line illustrates the direction in which the kernel $\dot{\phi}_t$ with parameters $\theta_t$ is learned or computed. 

For most generative models as in \Cref{fig:TraditionalLearning} the path $S_t(\phi)$ is not known in a truly closed, analytically accessible form but is left open or is regressed from data using score-, flow- or Kullback--Leibler divergence oriented objectives~\cite{Liu:2016svg,Rezende:2016vinf,Dinh:2017nfd,Albergo:2019eim,Nicoli:2020njz,Ho:2020ddpm,Song:2021sbgm,Wang:2023exq,Lipman:2023fmgm,Xu:2024jko}. The parameters $\theta_t$ of the kernel at different times $t$ then usually depend on each other, and a change at one time affects the kernel at other times.

For PIKs in \Cref{fig:PIKLearning} the path $S_t(\phi)$ is given in a fully determined and accessible form. Then, the PIK setup allows one to read the Wegner or continuity equation \labelcref{eq:WF-dotphi} as a simple linear first order differential equation for the kernel $\dot{\phi}_t$. This effectively splits up the global (and possibly underdetermined) task from \Cref{fig:TraditionalLearning} of finding a generative map between two distributions into a set of small fully determined and independent tasks that allow for an error correction and systematic improvement. As illustrated in \Cref{fig:PIKLearning}, this also leads to kernel parameters $\theta_t$ at different times that are independent of each other.

Below, we highlight three key properties of the PIK architecture that follow from this structure and discuss them in the subsequent sections.
\begin{enumerate} 
\item[(i)] \textit{Independent kernels}: PIKs convert the global task of finding a precise invertible map from the base distribution $p_0(t)$ to the final distribution $p_1(t)$ within a deep network into independent tasks of finding infinitesimal maps that solve for $\dot\phi_t(\phi;F_t)$ at each time step. For the pair $(S_t,\dot\phi_t)$ this amounts to solving \labelcref{eq:WF-dotphi} for $\dot{\phi}_t$. 
\item[(ii)] \textit{OOD resolution}: \Cref{eq:WF-dotphi,eq:GenRG} and respective equations for $F_t$ provide us with an error control of this map. They are also at the root of systematic improvements in the sampling process beyond the initial training time. In particular, curing the OOD problem with the PIRG pair $(S_t,\dot\phi_t)$ boils down to locally improving the solution $\dot \phi_t(\phi)$ to a known linear differential equation, whose coefficients are determined by $F_t(\phi)$.

\item[(iii)] \textit{Optimisation}: The properties (i) and (ii) are relevant for the setup and solution of PIKs for a given pair ($F_t, \dot{\phi}_t)$. The independent and analytically given structure of \labelcref{eq:WF-dotphi} offers further optimisation freedom for the pair, given external constraints. In particular, this includes a new kind of parameter conditional flow, systematic ways to improve inaccurate solutions of \labelcref{eq:WF-dotphi}, and the choice of the used generating function $F_t$. This is discussed in more detail in \Cref{sec:Optimisation}.
\end{enumerate} 
This closes the derivation and discussion of the general PIK architecture based on PIRG pairs $(F_t,\dot\phi_t)$. In the following we concentrate on the example pair $(S_t,\dot\phi_t)$ and discuss strategies to parametrise and optimise $S_t(\phi)$ and $\dot{\phi}_t(\phi)$. The respective findings also apply to general pairs.

\subsection{Sampling with the PIRG pair}
\label{sec:GenParaSt+dotphi}

Property (i) \textit{Independent kernels} leaves us with the task of devising solution strategies for single sampling steps at times $t$ with a pair $\bigl(F_t(\phi), \dot\phi_t(\phi)\bigr)$. The general case involves combinations of given $F_t$ and/or $\dot\phi_t$ and this generality is specifically important for the property (iii) \textit{Optimisation} of the PIK architecture. Here, we concentrate on solution strategies of \labelcref{eq:WF-dotphi} with fixed $S_t$.

\subsubsection{Parametrisation of $S_t(\phi)$}
\label{sec:GenParaSt}

In this case, we must choose a suitable parametrisation of $S_t(\phi)$ and following \labelcref{eq:GenDistr}, we choose 
\begin{subequations} 
\label{eq:S-likelihood} 
\begin{align}
    S_t(\phi) = \hat{S}_t(\phi) + \log {\cal N}_t\,.
    \label{eq:lSikelihood}
\end{align}
Here, $\hat{S}_t(\phi)$ contains all field-dependent parts and $\log {\cal N}_t$ comprises a field-independent contribution that ensures the normalisation. We parametrise
\begin{align}
    \hat{S}_t(\phi) = \sum_i c_{i, t} \, \mathcal{O}_i(\phi)\,,
    \label{eq:ActionParametrization}
\end{align}
\end{subequations}
with scale-dependent coefficients $c_{i, t}$ and a set of analytically known basis functions $\mathcal{O}_{i,t}(\phi)$. The basis functions can be chosen $t$-independently but in specific cases $t$-dependent ones may be advantageous for covering non-linear transformations within a smaller set of operators, see e.g.~\cite{Lamprecht2007,Isaule:2018mxt,Isaule:2019pcm}. At the initial and final scale, $\{c_{i}\,,\, \mathcal{O}_{i} \}$ are chosen such that \labelcref{eq:lSikelihood} with \labelcref{eq:ActionParametrization} matches the known base and target negative log likelihoods $S_0, S_1$. In between, the only constraint for the trajectory is its differentiability, and that \labelcref{eq:S-likelihood} allows for an analytical computation of $\partial S_t(\phi)/\partial \phi$. However, secondary constraints can be used for the optimisation of the sampling process. Furthermore, this also opens the opportunity to learn optimal base distributions within the PIK architecture.

The final important ingredient in the PIK architecture is the normalisation of the flowing distribution $p_t$, also present in the normalisation flow architecture. For PIKs the flow of the normalisation ${\cal N}_t$ is directly related to that of $S_t$: with \labelcref{eq:dtZ0} and \labelcref{eq:S-likelihood} the change of the normalisation ${\cal N}_t$ can be computed from that of $\hat S_t$, 
\begin{align}
    \frac{d \log {\cal N}_t}{dt} =  - \int {\cal D} \phi\,p_t(\phi) \frac{d\hat{S}_t(\phi)}{dt} \,.
    \label{eq:ChangeInPartitionSum}
\end{align}
The explicit resolution of \labelcref{eq:ChangeInPartitionSum} requires the computation of an expectation value for each 
$t$ from a non-trivial distribution $p_t(\phi)$. This would inflict large computational costs and in \Cref{sec:Optimisation} we show how this task can be converted into one of solving differences of Wegner equations, see the discussion around \labelcref{eq:OptimizationCondition}. 

We close the discussion of the parametrisation of $S_t(\phi)$ with iterating the remark, that the analytic parametrisation of $S_t$ is but one of many possibilities. We have introduced it here and will use it later in the explicit examples in \Cref{sec:PIKsAtWork} for its concise and read-to-use structure.

\subsubsection{Parametrisation of $\dot{\phi}_t(\phi)$}
\label{sec:GenParadotphi}

The above parametrisation of $S_t$ fixes all $S_t$-dependent terms in the Wegner equation \labelcref{eq:WF-dotphi}, and leaves us with the task of solving it for the physics-informed kernel $\dot{\phi}_t(\phi)$. To solve \labelcref{eq:WF-dotphi} in practice, most known approaches require a parametrisation of the kernel $\dot{\phi}_t(\phi)$ in some set of basis functions $K_{j,t}(\phi)$, 
\begin{align}
    \dot{\phi}_t(\phi) = \sum_j k_{j, t} \, K_{j,t}(\phi)\,,
    \label{eq:KernelParametrization}
\end{align}
with a chosen complete basis $\{K_{j,t}(\phi)\}$ and expansion coefficients $k_{j, t}$. With \labelcref{eq:WF-dotphi}, this reduces the complicated optimisation problem of finding a transformation between two distributions into the problem of solving an independent set of linear systems of the form
\begin{align}
    A_t\, k_t = b_t\,.
    \label{eq:LinearSystem}
\end{align}
Here, the matrix $A_t$ and vector $b_t$ are directly determined by the chosen basis $K_{j,t}(\phi)$ and the parametrised $S_t(\phi)$. Note, that because \labelcref{eq:LinearSystem} is linear, finding the expansion coefficients $k_{j,t}$ directly avoids problems related to local minima.

\subsection{Learning and optimisation}
\label{sec:Optimisation}

The optimisation of the PIK architecture amounts to optimising the global solution for $\dot{\phi}_t$, that is that of \labelcref{eq:LinearSystem} for all $t$. In this process we aim for both the most rapid convergence of the given expansion and the stability of the solution for $\dot{\phi}_t$. This is in one-to-one correspondence with, firstly, a local and global optimisation of the basis $K_{j,t}(\phi)$ and, secondly, an optimisation of the determination of the respective coefficients $k_{j, t}$. This local optimisation of the basis and the coefficients is addressed in \Cref{sec:OptimisationSolution}. Moreover, the global optimisation of the pair $(F_t, \dot{\phi}_t)$ is addressed in \Cref{sec:OptimisationPair}.

\subsubsection{Local optimisation}
\label{sec:OptimisationSolution}

In the PIK architecture, the problem of finding maps between two distributions is reduced to solving \labelcref{eq:WF-dotphi} or rather \labelcref{eq:LinearSystem} for each layer. This entails that the local optimisation task in PIKs does not require machine learning. Practically speaking, the solution of \labelcref{eq:WF-dotphi,eq:LinearSystem} thrives on any successful method to solve high-dimensional linear PDEs; this includes traditional as well as machine learning-based methods. Accordingly, PIKs may immensely benefit from the extensive optimisation framework to improve traditional methods as well as neural network-based approaches to solve high-dimensional PDEs such as apt physics-informed neural networks or operators~\cite{Rassi:2019zul, Li:2023pino, Richter-Powell:2022xww}. Moreover, recent approaches to learn optimal basis functions for PDE solvers may significantly improve the solutions computed with PIKs~\cite{Weng:2025dcm}.

In practice, both traditional and machine learning-based solution strategies have to deal with the fact that \labelcref{eq:WF-dotphi} depends on the change of the normalisation which is not known analytically: $d \log {\cal N}_t/dt$ in \labelcref{eq:ChangeInPartitionSum} is only determined by an expectation value. However, we can use the fact that $d \log {\cal N}_t/dt$ is a field-independent constant to evade its explicit computation. To that end, we consider differences of Wegner equations \labelcref{eq:WF-dotphi} evaluated at the targeted field configuration $\phi$ and at some reference configuration $\chi$. Since $d \log {\cal N}_t/dt$ is field-independent, it drops out in the difference. Thereby we circumvent the explicit computation or equivalently learning of $d \log {\cal N}_t/dt$ as done in related approaches~\cite{Mate:2023fxa,Albergo:2025nets}. This leaves us with an efficient condition that holds for any allowed pair $(S_t, \dot{\phi}_t)$,
\begin{align}\nonumber 
    0 =  \frac{d \hat{S}_t(\phi)}{dt} &- \frac{d \hat{S}_t(\chi)}{dt} \\[1ex]
    &- \left[\,\frac{\partial}{\partial \phi} - \frac{\partial \hat{S}_t(\phi)}{\partial \phi} \, \right] \, \dot{\phi}_t(\phi) \nonumber \\[1ex] 
    &+ \left[\frac{\partial}{\partial \phi} - \frac{\partial \hat{S}_t(\phi)}{\partial \phi} \,\right] \, \dot{\phi}_t(\phi)\, \Big|_{\phi=\chi}\,. 
    \label{eq:OptimizationCondition}
\end{align}
Crucially, this condition only requires terms that are analytically known or efficiently computable.

\subsubsection{Global optimisation}
\label{sec:OptimisationPair}

As mentioned before, besides optimising the solution process of \labelcref{eq:WF-dotphi}, the PIK architecture also allows to optimise the pair $(F_t, \dot{\phi}_t)$ with respect to some secondary criterion. Here, we aim to highlight a list of said criteria and comment on how they may be implemented.

Firstly, in many applications one is not merely interested in sampling from one specific distribution, but rather a family of distributions that depend on some external parameters~\cite{Zhu:2024kiu,Winkler:2023lcf,Ardizzone:2019fwy,Gerdes:2022eve,Singha:2022icw,Wahl:2024trade}. To that end, parameter-conditional generative models were devised. The majority of these models work by enlarging the generative model and learning the additional parameter dependence. The parameter conditional simulations are then conducted by multiple forward passes through the enlarged network. However, for PIKs we have a direct control over the distributions seen during one forward pass, making the additional simulations and enlarged network superfluous. This will be discussed in more detail in \Cref{sec:Optimisation_0D}. Note that in principle this procedure could also be adapted for stochastic normalising flows and non-equilibrium transport sampler~\cite{Wu:2020snf,Albergo:2025nets}.

Secondly, while the parameter-conditional mode of PIKs is an interesting choice, it requires a precise solution of each intermediate kernel $\dot{\phi}_t$ during the flow. However, if one is only interested in the final distribution, one can optimise the pair ($F_t, \dot{\phi}_t$) in different ways. For instance, one may absorb the inaccuracies of the intermediate kernels $\dot{\phi}_t$ into the path $F_t$, requiring only a precise solution for the final kernel. Because we have analytical access to \labelcref{eq:OptimizationCondition}, we can do so in a controlled and possibly iterative manner. 

This goes hand in hand with optimising the base distribution $p_{t=0}$. The latter offers a lot of potential for optimisation, as its only requirement is that one should be able to sample from it efficiently~\cite{Bauer:2024byr}. Moreover, sampling from the best base distribution for a given transformation is known from trivialising flows~\cite{Luscher:2009eq,Albandea:2023wgd,Albandea:2023ais,Foreman:2021ljl}. The \textit{independent kernel} property and \labelcref{eq:WF-dotphi} may allow one to determine closed forms for these base distributions for a given transformation, that were not within reach before.

Thirdly, while most comments above were made with the action (or negative log likelihood) in mind, we readily add that PIKs can also offer extensions to other generating functions $F_t$ such as the rate function that store the information of the distribution in a more condensed manner. This is part of future work. 

Lastly, the above comments only related to optimising the pair $(F_t, \dot{\phi}_t$) for generative tasks. However, PIKs can, in principle, also be used to infer analytical forms of the action or Hamiltonian from a given data set. This allows to interpret the physics underlying the data set and connects PIKs to symbolic regression tasks. Practically, it is analogous to approaches based on contrastive divergence~\cite{Hinton:2002tpe}, where one considers whether a Langevin process with the drift of the action/Hamiltonian in question moves away from the empirical distribution given by the data. Similar to approaches based on Stein's identity~\cite{Liu:2016svg,Grathwohl:2020fzl}, with PIKs one can replace the Langevin process with the flow induced by the kernel $\dot{\phi}_t$. Then, optimising the kernel such that it corresponds to a trivial transformation allows to infer and optimise the analytical form of the action belonging to the data set.\vspace{2mm}

This closes the derivation of the PIK architecture and its general solution procedure. In the following sections we illustrate a concrete implementation and its strengths at the example of scalar lattice field theories.

\section{PIKs at Work}
\label{sec:PIKsAtWork}

We proceed with selected applications of the PIK architecture to tailor-made examples: scalar field theories on the lattice. These examples allow us to illustrate many of the advantageous properties of PIKs in a well-controlled setting. This Section is meant to practically illustrate the considerations made in \Cref{sec:Introduction,sec:PIKs}. In \Cref{sec:Introduction}, we considered two OOD problems for generative models. 

The first OOD problem relates to the size of the system: Transformations computed for a high-dimensional problem are typically inaccurate as they must interpolate in a high-dimensional space. 
The second OOD problem relates to the size of the training data: Transformations computed on a limited data set cannot generally be expected to have obtained knowledge on cumulants that are not sufficiently represented in the training data.

In \Cref{sec:phi4-0D}, we show how PIKs systematically resolve the second OOD problem: Since PIKs transform the generative task into the task of solving a sequence of independent, linear PDEs and solutions to linear PDEs can be systematically improved, PIKs inherit the same property. Accordingly, one can provide PIKs with information that goes beyond the initial training data in a controlled manner.

To that end, we begin by showing how one can construct PIKs that benefit from their (i) \textit{independent kernels} at the example of a zero-dimensional $\phi^4$-theory. Then, we explicitly comment on PIKs (ii) \textit{OOD resolution} by correcting the PIK sampling process in a controlled way after its initial setup. Moreover, we illustrate how one can (iii) \textit{optimise} the PIRG pair $(S_t, \dot{\phi}_t)$ at the example of a parameter-conditional flow.

In \Cref{sec:phi4-dD} we illustrate the applicability of PIKs to higher-dimensional probability measures. We discuss, how the respective first OOD problem, the curse of dimensionality, reduces to that of solving high-dimensional analytically known linear PDEs whose form and sparseness can be optimised to the task at hand. The resolution of this OOD task is left to a forthcoming publication.

\subsection{Zero-dimensional $\phi^4$-theory}
\label{sec:phi4-0D}

As a first benchmark, we consider a one dimensional probability distribution $p(\varphi)= \exp\{-S(\varphi)\} $ with $\varphi \in \mathbbm{R}$ in \labelcref{eq:GenDistr}. In this simple case, \labelcref{eq:WF-dotphi} reduces to an ordinary differential equation. This example can be understood as a zero-dimensional scalar field theory and we restrict ourselves to a $\phi^4$-theory with the action 
\begin{align} 
\hat S(\varphi) =\frac12 m^2 \varphi^2 +\frac{\lambda}{4!} \varphi^4\,, \qquad \varphi \in \mathds{R}\,,
\label{eq:Svarphi0d}
\end{align} 
with the two parameters $m^2$ (mass squared) and $\lambda$ (coupling). This system can also be solved by direct integration which allows for a good comparison. 

For the resolution of the sampling problem with PIKs we define a PIRG pair with an action $S_t$, that is a linear interpolation between the base action $S_0$ and the target action $S_t$, similar to~\cite{Wu:2020snf}:
\begin{subequations} 
	\label{eq:Std=0}
\begin{align}
    \hat{S}_t(\phi) = \frac{1}{2} m^2(t) \,\phi^2 + \frac{\lambda(t)}{4}\, \phi^4\,, \qquad \phi \in \mathds{R}\,,
    \label{eq:ZeroDimensionalAction}
\end{align}
with 
\begin{align}\nonumber 
		m^2(t) =&\, m_0^2 + t\,\left( m_1^2 - m_0^2\right) \\
		\lambda(t) = &\, \lambda_0 + t\,\left( \lambda_1 - \lambda_0\right)\,. 
	\label{eq:CouplingParametrisation}
\end{align}
\end{subequations} 
Here $(m_0^2\, , \lambda_0)$ are the parameters of the base distributions and $(m_1^2\, ,\lambda_1)$ are those of the target distribution. Note that this parametrisation is a free choice and naturally lends itself to parameter-conditional models as will be discussed in \Cref{sec:Optimisation_0D}. This is a specific kind of optimisation and other examples can be found in \Cref{sec:Optimisation}.

For this toy model, we use a flow from a unimodal Gaussian theory with 
\begin{align} 
m_0^2 = 1  \,, \quad \lambda_0 =0\,,
\end{align} 	
to a theory with a multimodal behaviour and $\lambda\neq 0$, 
\begin{align} 
	m_1^2 = -2  \,, \quad \lambda_1 =\frac12\,. 
\end{align} 	
Note that capturing multimodal target distributions poses a difficulty for many traditional generative models because of the phenomenon of mode-collapse or mode-seeking~\cite{Hackett:2021idh,Nicoli:2023qsl,Chen:2022ytr}. PIKs circumvent this problem by enforcing a particular path $S_t$, which fixes the way the unimodal distribution is morphed into the multimodal one. Moreover, the independent kernels $\dot\phi_t$ allow for an error correction at each time step.

\subsubsection{Independent kernels $\dot\phi_t$}
\label{sec:IndependentKernels_0D}

During the flow, the field $\phi_t$ is propagated in time according to \labelcref{eq:FieldODE}. This ODE in time is solved numerically by a standard fourth order Runge-Kutta scheme with a step size of $\Delta t = 1 / 25$, leaving us with $N_t=51$ kernels $\dot{\phi}_t$ to be determined. Because we have fixed the path $S_t(\phi)$ via \labelcref{eq:ZeroDimensionalAction,eq:CouplingParametrisation}, each kernel has its own independent Wegner equation \labelcref{eq:WF-dotphi} it must solve.

In analogy to the parametrisation in \labelcref{eq:KernelParametrization}, we express each kernel $\dot{\phi}_t(\phi)$ in a chosen basis following a similar approach as in~\cite{Gerdes:2022eve},
\begin{align}\label{eq:SpecBasis}
    K_{1,t}(\phi) = \phi\,, \quad K_{j,t}(\phi) = \sin\left( \omega_{j,t} \,\phi \right)\quad j \geq 2\,.
\end{align}
We choose this basis because $\hat{S}_t(\phi)$ is symmetric under the transformation $\phi \to -\phi$. To ensure that the distribution generated by the PIKs carries the same symmetry, we want $\dot{\phi}_t$ to transform equivariantly under this sign flip~\cite{Koehler:2020efe}. 

In analogy to other generative models, we use stochastic gradient descent with the Adam optimiser to compute the coefficients $k_{j,t}$ and frequencies $\omega_{j,t}$. Furthermore, we use the mean squared error induced by \labelcref{eq:OptimizationCondition} as a loss function. Here, the fact that the change in normalisation $d \log {\cal N}_t/dt$ does not have to be computed is used explicitly. We choose $\chi=0$ as a reference configuration for \labelcref{eq:OptimizationCondition}, which leads to a vanishing time and field derivative of the action in \labelcref{eq:OptimizationCondition}. 

The loss is bounded from above by $10^{-4}$ after saturation. Accordingly, the effective sample size also effectively remains at unity. More details of the training are provided in \Cref{sec:TrainingDetails}.

\begin{figure}
	\centering
	\includegraphics[width=0.5\textwidth]{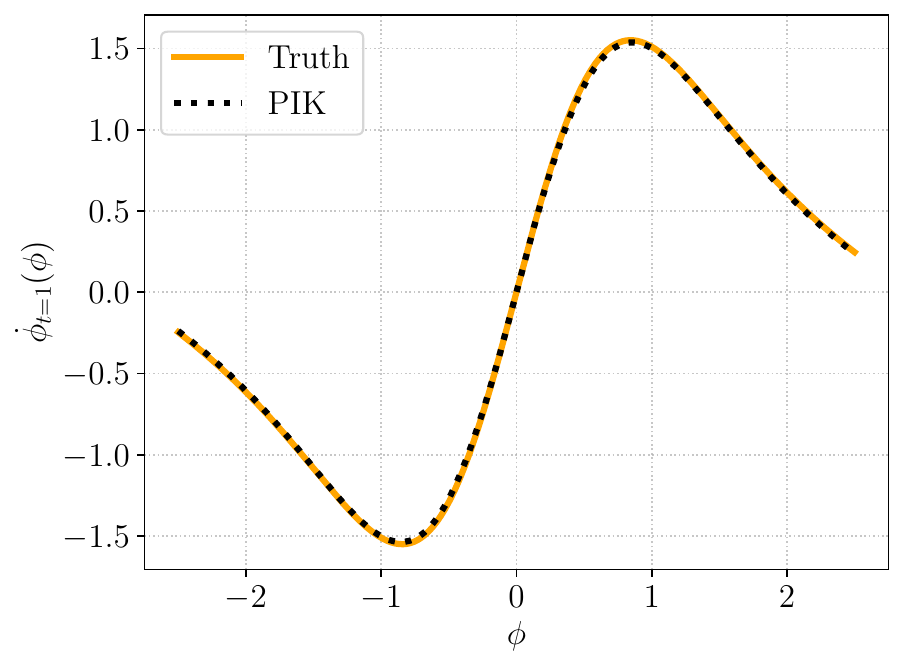}
	\caption{Comparison of the true (orange) and the learned (black) kernel $\dot{\phi}_{t=1}(\phi)$ for the zero-dimensional $\phi^4$-theory.  \hspace*{\fill} }
	\label{fig:PIKKernelTrueLearned}
\end{figure}

In order to show the successful training of the kernel, we compare the learned kernel $\dot{\phi}_t(\phi)$ at $t=1$ to the one obtained by directly integrating \labelcref{eq:WF-dotphi} in \Cref{fig:PIKKernelTrueLearned} and we find a very good overlap. 

\begin{figure}[t]
	\centering
	\includegraphics[width=0.5\textwidth]{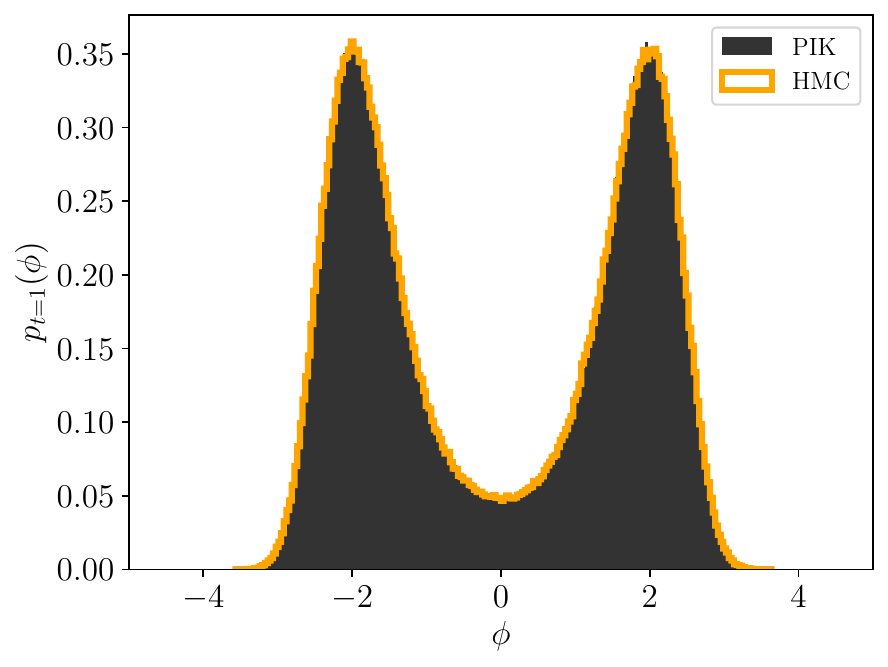}
	\caption{Comparison of the targeted (orange, HMC) and modelled (black, PIK) zero-dimensional $\phi^4$ distribution.  \hspace*{\fill} }
	\label{fig:PIKHMCDistribution_ZeroDimensional}
\end{figure}

\Cref{fig:PIKHMCDistribution_ZeroDimensional} shows a comparison of the targeted (orange) and modelled (black) distribution at the final time $t=1$. We find that the well-trained PIK transformation indeed connects the initial Gaussian distribution to the targeted multimodal one. The target distribution was computed using $10^6$ samples from a standard hybrid Monte Carlo algorithm (HMC). The modelled distribution was generated by pushing $10^6$ samples from the initial Gaussian distribution through the PIK flow. We find an excellent agreement between both distributions.

We close the discussion of the independent kernels $\dot\phi_t$ with a hands-on assessment of the speedup in runtime gained by choosing PIKs over HMC. Evidently, the performance of both HMC and PIKs depends crucially on the specific implementation~\cite{Ostmeyer:2024amv}, and we refrain from optimising neither HMC nor PIKs for speed. Such a comprehensive analysis goes beyond the scope of the present work, and hence the present comparison only offers a rough estimate for the speedup. Specifically, for HMC we use the standard leapfrog integrator with the same discretisation in time as for PIKs and integrate trajectories up to a final time $T=1$. The number of effectively independent samples generated by HMC is computed using the $\Gamma$-method~\cite{Wolff:2003sm}. With this setup, one generates the same number of independent samples with PIKs around $20$ times faster compared to HMC. While only being a rough estimate, we consider such a speedup very promising. 

%

\subsubsection{OOD resolution}
\label{sec:OOD-resolution_0D}

Having showcased that PIKs have the capacity to learn each kernel $\dot{\phi}_t$ independently without the requirement of computing the change in the normalisation, we now turn to the property (ii) \textit{OOD resolution}. Instead of the explicit volume scaling, here we illustrate the systematic and controlled improvability of the PIK sampling process after its initial setup and training. This is naturally enabled by the fact that we have analytical access to the Wegner equation \labelcref{eq:WF-dotphi} (or \labelcref{eq:OptimizationCondition}, respectively) that has to be satisfied by each kernel individually. It is further boosted by the fact that the Wegner equation is linear in $\dot{\phi}_t$, making the improvements more directly accessible.

As discussed in \Cref{sec:Introduction}, already the necessarily limited training data often leads to OOD problems for generative models. Here, one can use additional Monte Carlo steps to cure these problems~\cite{Wu:2020snf,Zhu:2025pmw}. However, since this can become costly and does not improve the inherent quality of the model, one may look for more optimal solutions. With PIKs, we can directly use the information of the distribution via $S_t(\phi)$ at each time step to systematically improve the sampling beyond training time.

\begin{figure}[t]
	\centering
	\includegraphics[width=0.5\textwidth]{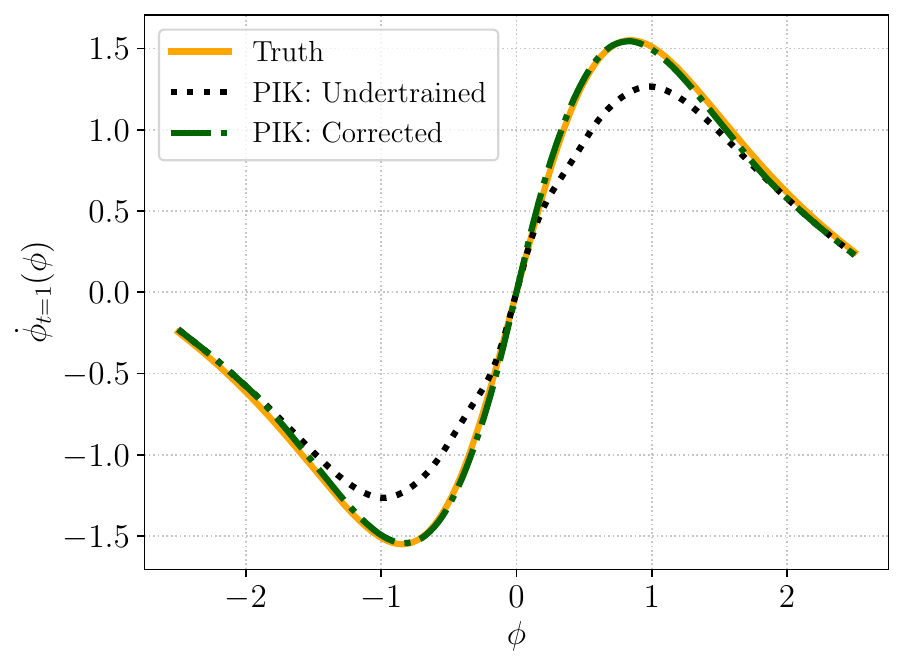}
	\caption{Comparison of the true (orange), the undertrained (black), and the corrected (green) kernel $\dot{\phi}_t(\phi)$ at $t=1$ for the zero-dimensional $\phi^4$-theory.  \hspace*{\fill} }
	\label{fig:PIKKernelTrueLearnedWithCorrection}
\end{figure}

In order to mimic this OOD problem, we purposefully undertrain the last layer of the PIK flow, such that the kernel only saw $\sim 50\,k$ samples during the optimisation. This undertrained or approximate kernel is referred to as $\dot{\phi}^{(0)}_t$. Note that also for other continuous generative models, the last layer must exactly solve \labelcref{eq:OptimizationCondition} in order to be truly generative. The undertrained kernel together with the true solution is depicted in \Cref{fig:PIKKernelTrueLearnedWithCorrection} in black and orange, respectively. Here, a clear deviation between the two is visible.

In order to systematically improve the sampling, we must enrich the solution of the kernel $\dot{\phi}^{(0)}_t$ such that it better satisfies \labelcref{eq:OptimizationCondition}. There are manifold ways to achieve this in a controlled way~\cite{Cavoretto:2020art,Efendiev:2013gma,Chung:2018gqo,Weng:2025dcm,Schaback:2000agt,DeVore:1996gag}. Here, we use the linear structure of $\labelcref{eq:OptimizationCondition}$. For notational convenience, we denote the second and third line of \labelcref{eq:OptimizationCondition} as
\begin{align}
    \label{eq:LinearOperator_OptimisationCondition}
    \mathcal{L}_\chi \, \dot{\phi}_t(\phi) := &\left[\frac{\partial}{\partial \phi} - \frac{\partial \hat{S}_t(\phi)}{\partial \phi}\, \right] \, \dot{\phi}_t(\phi) \\[1ex]
    - &\left[\frac{\partial \hat{S}_t(\phi)}{\partial \phi} - \frac{\partial}{\partial \phi} \, \right] \, \dot{\phi}_t(\phi)\, \Big|_{\phi=\chi} \nonumber\,,
\end{align}
defining the linear operator $\mathcal{L}_\chi$ with the reference configuration $\chi$. The residual of the approximate kernel $\dot{\phi}^{(0)}_t$ is then readily computed as
\begin{align}
    r^{(0)}_\chi(\phi) = \frac{d \hat{S}_t(\phi)}{dt} - \frac{d \hat{S}_t(\chi)}{dt} - \mathcal{L}_\chi \, \dot{\phi}^{(0)}_t(\phi)\,.
    \label{eq:Residual}
\end{align}
Naturally, a correction $\Delta \dot{\phi}^{(0)}_t(\phi)$ that satisfies
\begin{align}
    \mathcal{L}_\chi \, \Delta \dot{\phi}^{(0)}_t(\phi) = r^{(0)}_\chi(\phi)\,,
    \label{eq:CorrectionEquation}
\end{align}
will lead to a kernel
\begin{align}
    \dot{\phi}^{(1)}_t(\phi) = \dot{\phi}^{(0)}_t(\phi) + \Delta \dot{\phi}^{(0)}_t(\phi)\,,
\end{align}
that solves \labelcref{eq:OptimizationCondition} accurately. As is already hinted at by the superscript notation, this procedure can in principle be iterated because of the linear structure of the Wegner equation, allowing for controlled corrections~\cite{DeVore:1996gag}.

In order not to spoil the well-fitted regions of the undertrained kernel and because they more readily transfer to higher-dimensional problems, we solve \labelcref{eq:CorrectionEquation} with a collocation approach using compactly supported radial basis functions, more concretely Wendland functions~\cite{Wendland:1995ppp,Kansa:1990cma,Kansa:1990cmb}. To ensure the $\mathds{Z}_2$ symmetry of the transformation, the basis functions were anti-symmetrised. The centres of the radial basis functions were chosen from regions where the undertrained kernel deviated most from the true one. Here, we used eleven radial basis functions. Because the latter are compactly supported, the resultant linear system similar to \labelcref{eq:KernelParametrization} is sparse and readily solved. Further details of the correction procedure are provided in \Cref{sec:CollocationDetails}.

In \Cref{fig:PIKKernelTrueLearnedWithCorrection}, we show the corrected kernel $\dot{\phi}^{(1)}_t(\phi)$ in green together with the undertrained and true kernel in black and orange, respectively. We find that the correction indeed leads to a kernel that is very close to the true one and corrects the undertrained one in the regions where it deviates the most.

While the correction mechanisms can be readily improved, this example illustrates the general systematic workflow of PIKs from (a) an (imprecise) global solution to (b) a determination of OOD regions during sampling or using information of the residual to (c) a controlled (and local) correction of the kernel. This is natural for PDE approaches~\cite{DeVore:1996gag} but the proposed method is to our knowledge novel in the context of generative models.

\begin{figure}[t]
	\centering
	\includegraphics[width=0.5\textwidth]{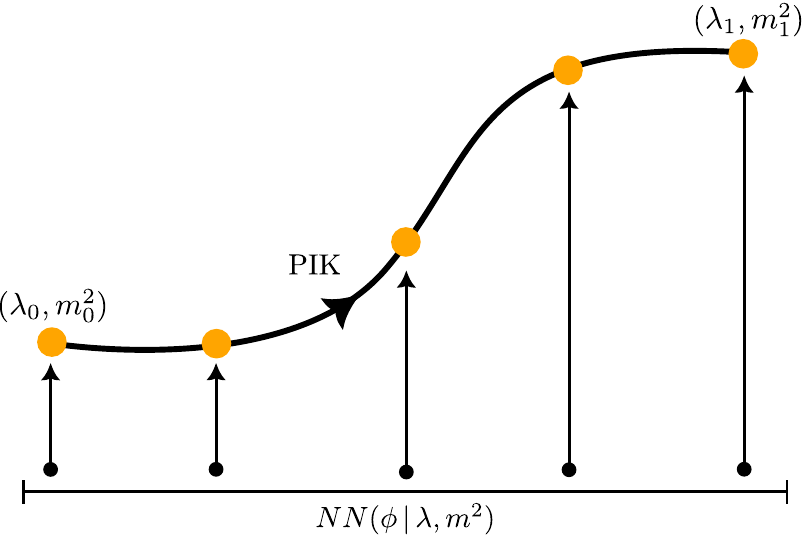}
	\caption{Difference in the kind of parameter conditionality for PIKs differs and traditional generative models $NN(\phi\,|\,\lambda, m^2)$. While PIKs move along the path determined by the parameters, traditional models need to sample each set of configurations individually.\hspace*{\fill}}
	\label{fig:ConceptualCouplingConditional}
\end{figure}
%

\subsubsection{Optimisation with parameter-conditional PIKs}
\label{sec:Optimisation_0D}

The above sections showcased the setup, training, and systematic improvement of PIKs. In this section, we comment on the property \textit{(iii) Optimisation} of PIKs. As outlined in \Cref{sec:Optimisation}, there are manifold ways to optimise the pair $(S_t, \dot{\phi}_t)$ in order to improve the sampling process, given auxiliary constraints. Here, we illustrate one such optimisation at the example of a parameter-conditional flow.

With PIKs we have direct control over the concrete distribution at each time step. Here, we use this freedom to enforce the same functional form of the action at each time step. The only thing that changes for the distribution in the flow are the parameters ($m^2, \lambda$) at each time step. This allows us to make PIKs parameter conditional without inflating the network. 

Unless one salvages a specific structure of the action~\cite{Zhu:2024kiu}, parameter-conditional networks $NN(\phi\,|\lambda, m^2)$ commonly work by enlarging the network in a constructive way such that it can learn the additional parameter-dependence~\cite{Gerdes:2022eve,Singha:2022icw}. For PIKs, the flow itself can be made parameter-conditional as described above such that any additional structure and learning becomes superfluous. This is also illustrated in \Cref{fig:ConceptualCouplingConditional}. Moreover, this directly allows us to compute observables in a continuous way without sampling the configurations for each parameter individually.

\begin{figure}
	\centering
	\includegraphics[width=0.5\textwidth]{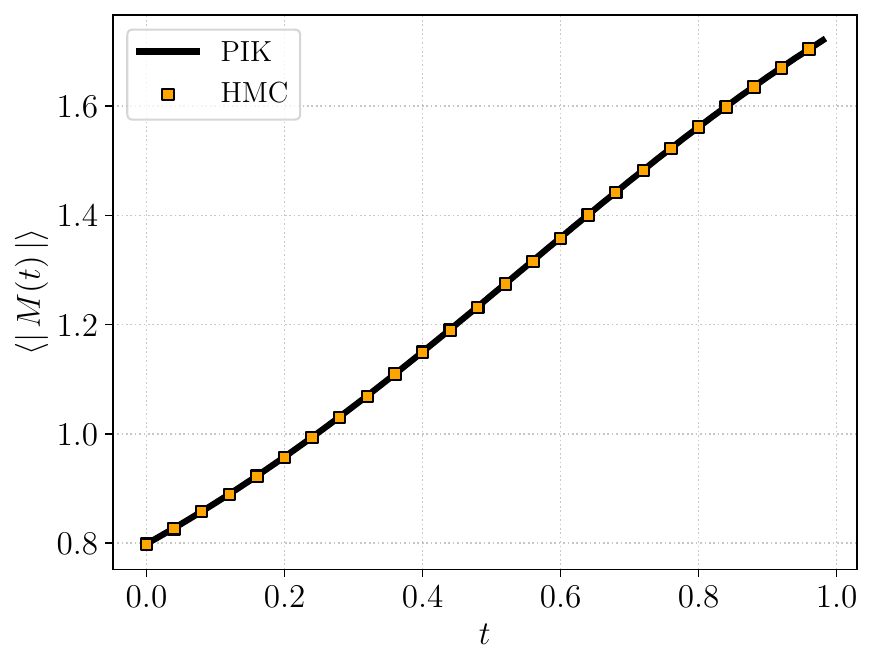}
	\caption{Magnetisation of the zero-dimensional $\phi^4$-theory computed with PIKs in a continuous manner (black) and HMC with individual simulations (orange). \hspace*{\fill}}
	\label{fig:CouplingConditional}
\end{figure}

We exemplify the PIK approach in \Cref{fig:CouplingConditional} for the observable
\begin{align}
    \langle | M(t) | \rangle = \int {\cal D}\phi \, \,p_t(\phi) \, |\phi|\,,
\end{align}
which corresponds to the expected magnetisation for the zero-dimensional $\phi^4$-theory. Here we draw a batch of samples from the Gaussian theory at $t=0$ and push it through the PIK flow. The observable (black) can then be computed at each intermediate time step and is purposefully depicted with a continuous line. For comparison, we also show the same observable computed with individual HMC simulations, where each point corresponds to a separate simulation.

\subsection{Higher-dimensional $\phi^4$-theories}
\label{sec:phi4-dD}

In the previous \Cref{sec:phi4-0D}, we have used the zero-dimensional $\phi^4$-theory to illustrate and benchmark PIKs in an analytically accessible example. In particular, we have showcased how PIKs can evade the OOD problem pertaining to the limited information in the training data. The remaining OOD problem, the curse of dimensionality, is only present in higher-dimensional systems and its proposed resolution is discussed here. In \Cref{eq:PIK-Highd} we discuss the general application of PIKs to higher-dimensional systems. In \Cref{sec:CurseDimensionality} we detail how PIKs can be used to systematically tackle the curse of dimensionality. A full numerical study goes beyond the scope of the present work and is the subject of a forthcoming publication.

As an exemplary system, we consider a $d$-dimensional lattice $\phi^4$-theory with $V$ sites and the action
\begin{align}\nonumber 
      \hat{S}_t(\phi) = &\,\sum_{\hat{n}\in \mathcal{D}} \Bigg[ \frac{1}{2} \left(m^2(t) + 2d\right) \phi_{\hat{n}}^2 \\[1ex]
      &\hspace{1cm}- \sum_{\mu=1}^d \phi^{\phantom{}}_{\hat{n}}\,\phi^{\phantom{}}_{\hat{n}+\hat{\mu}} + \frac{\lambda(t)}{4}\phi_{\hat{n}}^4 \Bigg]\,. 
 \label{eq:Phi4Action_dDimensional}
 \end{align}
Here, the coupling $\lambda(t)$ and squared mass $m^2(t)$ are linear in $t$ and are given by \labelcref{eq:CouplingParametrisation}. Moreover, $\hat{\mu}$ is the unit vector in the direction $\mu$ with $\mu=1,...,d$. Specifically interesting corners in these higher dimensional theories are the regimes that host second order phase transitions. There, standard algorithms face critical slowing down as the field values on well-separated lattice sites become strongly correlated which makes traditional sampling algorithms such as HMC inefficient~\cite{Wolff:1989wq}. Moreover, also known generative models struggle to learn adequate transformations in this regime~\cite{Abott:2022ass,DelDebbio:2021emt}. Below we discuss how PIKs can help to tackle these problems for generative models.

\subsubsection{PIK setup for high-dimensional theories}
\label{eq:PIK-Highd}

High-dimensional systems introduce additional redundancies, most importantly symmetries, that leave the action invariant. For generative models, these redundancies can be treated with equivariant flows in order to improve the sampling quality~\cite{Koehler:2020efe,Gerdes:2022eve,Boyda:2020hsi,Gerdes:2024rjk}. In the following, we show that also PIKs are able to easily and efficiently eliminate these redundancies. In doing so, we further show how solving for the kernel can be reduced to solving for a single function $\Omega_t$ instead of a vector field $\dot \phi_{t}$ directly, see also~\cite{Koehler:2020efe,Aarts:2025pmc,Albergo:2025nets}. This is accompanied by a discussion of boundary conditions, which implement the normalisation of the distribution and define a unique solution of the kernel.

As mentioned above, PIKs as well as standard generative approaches optimise a high-dimensional vector field, which entails optimising $V$ functions. This process can be simplified by writing the kernel as a gradient field~\cite{Koehler:2020efe,Aarts:2025pmc,Albergo:2025nets} and in the context of PIKs the kernel in the Wegner equation \labelcref{eq:WF-dotphi} admits a Helmholtz decomposition. Thus, we can express the kernel in terms of a function $\Omega_t(\phi)$ and a vector field $H_t(\phi)$
\begin{align}
    \label{eq:HelmholtzDecomposition}
    \dot{\phi}_t(\phi) &= -\frac{\partial}{\partial \phi}\, \Omega_t(\phi) + H_t(\phi)\,, \\[1ex]
    0 &= \frac{\partial}{\partial \phi}\,\left[\, p_t(\phi)\, H_t(\phi)\,\right]\,. \nonumber
\end{align}
For the Wegner equation, this directly manifests as 
\begin{align}
    \frac{d S_t(\phi)}{dt} = -&\left[\frac{\partial}{\partial \phi} - \frac{\partial S_t(\phi)}{\partial \phi}\, \right] \, \frac{\partial}{\partial \phi}\, \Omega_t(\phi) \\[1ex]
    + &\underbrace{\left[\,\frac{\partial}{\partial \phi} - \frac{\partial S_t(\phi)}{\partial \phi} \,\right] H_t(\phi)}_{=0}\,. \nonumber
\end{align}
Accordingly, the additional vector field $H_t(\phi)$ does not contribute to the change in the action. Rather, it describes permutations and rotations in the field space that are irrelevant for the change in the measure.
Typically, one would like to discard those to improve the sampling quality~\cite{Koehler:2020efe} and it is a nice feature of PIKs that it does so naturally.
With this, rather than optimising and parametrising $V$ functions, one must merely solve the adapted Wegner equation
\begin{align}
    \label{eq:WF-ScalarPotential}
    \left[\, \frac{\partial^2}{\partial \phi^2} - \frac{\partial S_t(\phi)}{\partial \phi}\, \frac{\partial}{\partial \phi}\,\right] \Omega_t(\phi) &= - \frac{d S_t(\phi)}{dt}\,,
\end{align}
for one function $\Omega_t(\phi)$. Here, $\partial^2/\partial \phi^2$ denotes the canonical Laplace operator. Note that \labelcref{eq:WF-ScalarPotential} is a linear PDE of elliptic type. Accordingly, it benefits from the insights gained in \Cref{sec:phi4-0D}.

To determine the suitable boundary conditions, we integrate the Wegner equation \labelcref{eq:WF-dotphi} over a hypercube $\mathcal{V}$ with radius $R$ in all directions. Using the divergence theorem, this straightforwardly leads to
\begin{align}
    - \int_\mathcal{V} \mathcal{D}\phi\, p_t(\phi) \, \frac{d \hat{S}_t(\phi)}{dt}  &- \frac{d \log \mathcal{N}_t}{dt} = \\[1ex]
    &\int_{\partial \mathcal{V}}\,\mathcal{D}\phi_s \,p_t(\phi_s) \, \dot{\phi}_t(\phi_s) \,. \nonumber
\end{align}
Here, the right-hand side denotes a surface integral over the boundary $\partial \mathcal{V}$ of the hypercube and $\mathcal{D}\phi_s$ refers to the respective measure. Letting $R \to \infty$, one may use the definition of $\log \mathcal{N}_t$ and \labelcref{eq:ChangeInPartitionSum} to see that the left-hand side vanishes. This leads to the boundary condition
\begin{align}
    \label{eq:BoundaryCondition}
    \lim_{R \to \infty} \int_{\partial \mathcal{V}}\,\mathcal{D}\phi_s \,p_t(\phi_s) \, \dot{\phi}_t(\phi_s) = 0\,,
\end{align}
for the kernel and ensures the normalisation of the distribution during the flow. 

In summary, \labelcref{eq:WF-ScalarPotential} together with \labelcref{eq:HelmholtzDecomposition,eq:BoundaryCondition} encodes a complete, redundancy-reduced version of the propagation step for the generative task with the path $S_t$.

\subsubsection{Tackling the curse of dimensionality}
\label{sec:CurseDimensionality}

In standard black-box approaches to the generative task, reducing the complexity of finding an accurate vector field is typically limited to salvaging known symmetries of the theory~\cite{Gerdes:2022eve,Koehler:2020efe,Kanwar:2020xzo}. In PIKs however, the curse of dimensionality can be tackled by using the analytic structure at hand to reduce the complexity of the system: the adapted Wegner equation \labelcref{eq:WF-ScalarPotential} provides us with insights about a good basis, leading to a more rapid convergence to the solution. For example, in the non-interacting case ($\lambda=0$) the Eigenfunctions of the operator on the left-hand side of \labelcref{eq:WF-ScalarPotential} are properly chosen tensor products of Hermite polynomials. Since Hermite polynomials can be used to form a complete basis, they are well-suited to expand $\Omega_t$. Compared to black-box approaches, this is assumed to reduce the number of required optimisable parameters.

For the present example \labelcref{eq:Phi4Action_dDimensional}, fully optimised PIKs may allow to limit the number of required basis elements to only grow with the spacetime dimension $d$ rather than the volume $V$. To that end, we note that the right-hand side of \labelcref{eq:WF-ScalarPotential} only includes terms $\phi_{\hat{n}}^2, \phi_{\hat{n}}^4$ that do not couple different lattice sites, see \labelcref{eq:Phi4Action_dDimensional}. Moreover, because of the local structure of $S_t$, the differential operator in \labelcref{eq:WF-ScalarPotential} maps tensor product of nearest neighbours back to tensor products of nearest neighbours. However, tensor products that include more distant lattice sites generate new terms that need to be cancelled again as they do not appear on the right-hand side. Note that different to CNNs, this parameter reduction does not only follow from the translational invariance of the system but rather from the local structure of the action and the Wegner equation, which is a stronger statement.

We may also utilise the property of the PIK, that the kernel transformation implicitly also includes a standard rescaling RG step. This property becomes even explicit if changing the generating function $F_t=S_t$ to the Wilson action or the effective action. This is analogous to the inverse RG as used in \cite{Bauer:2024byr,Bachtis:2021eww,Efthymiou:2019qnc,Singha:2025mgs,Masuki:2025irg} and will be discussed in a forthcoming work.

Finally we can optimise the adapted Wegner equation \labelcref{eq:WF-ScalarPotential} by varying  the path $S_t$. This changes the $S_t$-dependent coefficients of \labelcref{eq:WF-ScalarPotential}, see \Cref{sec:OptimisationPair}. Optimised choices for the coefficients simplify the task of solving \labelcref{eq:WF-ScalarPotential}. 

In summary, the solution of fully optimised PIKs may scale with the spacetime dimension as discussed above. This would reduce a problem with a na\"ive scaling with the volume to one that only scales with the spacetime dimension. Already without using this additional structure, promising results have been obtained in the related approach~\cite{Albergo:2025nets} within an application to a two-dimensional $\phi^4$-theory. Here, the PIK setup allows us to additionally exploit the structure provided by the Wegner equation. The remaining challenge of ensuring the accuracy of the solution to the Wegner equation for PIKs is then reduced to an OOD problem of the second type whose solution was already illustrated in \Cref{sec:OOD-resolution_0D}.

\section{Summary and outlook}
\label{sec:Outlook}

In this work, we introduced physics-informed kernels (PIKs) as a new architecture for generative models. The PIK-architecture aims at solving OOD problems in generative models as well as implementing a systematic error control. At its heart, PIKs reduce the task of finding transformations between two distributions into a decoupled chain of layerwise transformations that are given by the analytic continuity equations. For the distribution itself the continuity equation is the Wegner equation \labelcref{eq:Wegner74}, for the rate function the continuity equation is the generalised flow equation \labelcref{eq:GenRG}, and similar ones for other generating functions. Viewed as PDEs for the kernels these equations boil down to independent linear differential equations whose solution can be systematically corrected. As a paradigmatic example, we have considered PIKs for the distribution with completely determined action paths $S_t(\phi)$ that lead to simple Wegner equations. 

In general, PIKs thrive in situations where one has some access to the action, Hamiltonian, or negative log likelihood of the system at hand. In comparison to diffusion models, flow matching, and related architectures~\cite{Rezende:2016vinf,Dinh:2017nfd,Albergo:2019eim,Nicoli:2020njz,Ho:2020ddpm,Song:2021sbgm,Wang:2023exq,Lipman:2023fmgm,Xu:2024jko,Masuki:2025irg}, PIKs offer direct ways to implement this knowledge into the training and sampling without requiring Monte Carlo steps or simulations. In short, PIKs aim to use this property to systematically tackle OOD problems in generative architectures.  

In the present work we have shown how one can set up PIKs with (i) \textit{independent Kernels}. Moreover, we illustrated that one can implement an error correction to systematically approach a (ii) \textit{OOD resolution}. Lastly, we illustrated that one can use the gained freedom of PIKs to (iii) \textit{optimise} the PIRG pair $(S_t, \dot{\phi}_t)$ at the example of a new form of parameter-conditional generative models. 

Thus, this work paves the way for many interesting applications and improvements of PIKs highlighted throughout the text. 

The most prominent direction is the application of PIKs to currently studied theories in physics and related fields. This avenue includes the application to higher-dimensional theories as detailed in \Cref{sec:phi4-dD}. It further includes incorporating gauge and fermionic degrees of freedom into the flow. Since the respectively adapted Wegner equation is known, PIKs can be straightforwardly applied to these systems. In these applications to currently relevant theories, PIKs explicitly target OOD problems encountered for generative models and are therefore an interesting target to show the practical advantages of generative sampling.

Supporting these developments opens up further avenues for future research by the manifold opportunities for both local and global optimisation as detailed in \Cref{sec:Optimisation}. PIKs dissect the generative task into solving sets of linear differential equations. Systematically improving (parts of) this solution process is an important technical task ideally suit for machine learning applications. This includes salvaging PINNs, neural operators, optimised basis expansions, or improved learned paths for the action to solve the respective equations more efficiently. Targeting the global optimisation, opens up the possibility to learn optimal base distributions, benefit from improved generating functions $F_t$, or infer the action for a given data set, see \Cref{sec:OptimisationPair}.

In combination, PIKs form a generative model that splits the generative task into well-controlled sub-problems given by analytically known and independent differential equations determined by the underlying physics. This structure allows for systematic corrections, optimisations of individual parts of the algorithm, and thereby targets OOD problems in generative models.

\section*{Acknowledgments}

We thank Gert Aarts, Kenji Fukushima, Sander Hummerich, Thore Kolja Joswig, Ullrich Köthe, Timoteo Lee, Manfred Salmhofer, Robert Scheichl, Lingxiao Wang and Kai Zhou for discussions and collaborations on related subjects. FI and RK would like to thank the Isaac Newton Institute for Mathematical Sciences, Cambridge for support and hospitality during the programme RCLW04, supported by the EPSRC grant no EP/Z000580/1. This work is part of the DM-LFT collaboration. It is funded by the Deutsche Forschungsgemeinschaft (DFG, German Research Foundation) under Germany’s Excellence Strategy EXC 2181/1 - 390900948 (the Heidelberg STRUCTURES Excellence Cluster). We also acknowledge support by the
state of Baden-Württemberg through bwHPC.

\appendix

\section{Training details}
\label{sec:TrainingDetails}

In \Cref{sec:IndependentKernels_0D}, we have discussed how one can compute the independent kernel using standard machine learning techniques. In this Appendix we expand on the details of the present implementation.
As described in the introduction, at each time-step $t$ an independent kernel is computed using the samples $\phi_t$, which then provides the new $\phi_{t+\Delta t}$.
Samples for the training of the kernels are drawn from the base distribution defined by $S_0$ and successively evolved along the RG-time $t$, by pushing them through previously computed kernels. Accordingly, for the training only samples from the base distributions are required.

The independent kernels are parametrised by \labelcref{eq:KernelParametrization} and basis functions are given by \labelcref{eq:SpecBasis}, using a linear term and sine-functions with learnable frequencies $\omega_{j,t}$. The respective coefficients and frequencies are computed using the Adam optimiser, optimising with respect to the mean squared error induced by \labelcref{eq:OptimizationCondition}. 

To compute the RG-time evolution, we chose a time step $\Delta t = 1 /25$ within the fourth order Runge-Kutta scheme, which requires the determination $N_t = 51$ kernels. For each kernel, we chose $N_f=25$ learnable frequencies. The first two time-steps kernels were trained with a learning rate of $10^{-2}$ for 600 epochs using a batch size of $512$ samples that were drawn from the respective initial distributions. For the subsequent kernels, we transferred the weights of the previous kernel and trained them with a learning rate of $10^{-3}$ for maximally 600 epochs and used early stopping w.r.t the loss function to reduce the training time. For all kernels, this training procedure achieved a mean squared error loss that was bounded from above by $10^{-4}$ after training, achieving an effective sample size very close to unity.

After writing, we took note that this approach of learning the kernels iteratively is similar in spirit to an approach put forward in~\cite{Xu:2024jko}, where the map is regressed from data of an unknown distribution. Note however, that PIKs differ qualitatively from this approach as they use a completely known path $F_t$ for the generating function, transfer the generative task to solving linear differential equations, and allow for systematic corrections beyond the initial training data.

\section{Correction by collocation methods}
\label{sec:CollocationDetails}

In \Cref{sec:OOD-resolution_0D}, we discussed how one can systematically correct a kernel that captures some but not all of the relevant features of the true kernel. While there are many possible mechanisms to improve the solution of a PDE~\cite{Cavoretto:2020art,Efendiev:2013gma,Chung:2018gqo,Weng:2025dcm,Schaback:2000agt,DeVore:1996gag}, here we illustrate one such method based on a collocation approach with radial basis functions~\cite{Kansa:1990cma,Kansa:1990cmb}, which is a simple and straightforward choice.

As discussed in \Cref{sec:OOD-resolution_0D}, for an imperfect kernel $\dot{\phi}^{(0)}_t(\phi)$, one can easily compute the residual $r^{(0)}_\chi(\phi)$ w.r.t \labelcref{eq:OptimizationCondition} and the reference configuration $\chi$. To correct the kernel, we would like to solve the linear PDE
\begin{align}
    \mathcal{L}_\chi \, \Delta \dot{\phi}^{(0)}_t(\phi) = r^{(0)}_\chi(\phi)\,,
\end{align}
for the correction term $\Delta \dot{\phi}^{(0)}_t(\phi)$~\cite{DeVore:1996gag}. The linear operator $\mathcal{L}_\chi$ was defined in \labelcref{eq:LinearOperator_OptimisationCondition}. Similar to \labelcref{eq:KernelParametrization}, we parametrise $\Delta \dot{\phi}^{(0)}_t(\phi)$ as 
\begin{align}
    \Delta \dot{\phi}^{(0)}_t(\phi) = \sum_{i=1}^{M} w_i \, W_i(\phi, \phi_i)\,.
\end{align}
Here, $\{\phi_i\}_{i=1}^M$ denote the fixed centres of the radial basis functions. We choose them to be in regions with a large mismatch between the true and the approximate kernel. Moreover, $W_i(\phi, \phi_i)$ relates to Wendland functions centred at $\phi_i$, which is a compactly supported and positive definite radial basis function~\cite{Wendland:1995ppp}. This allows for the a sparse linear system and local corrections to the approximate kernel. For the computations of this paper, we chose four times differentiable Wendland functions
\begin{align}
    W(r) = 
    (1-r)_+^{l+2} \left[ (l^2 + 4l +3) r^2 + (3l + 6) r + 3 \right]\,, 
\end{align}
with $r \in \mathds{R}$ and $l = \lfloor d/2 \rfloor +3$. Here, $(1-r)_+ = \textrm{max}(1-r,0)$ ensures the compact support. To ensure that $\Delta \dot{\phi}^{(0)}_t(\phi)$ is also odd, we anti-symmetrised the basis functions accordingly
\begin{align}
    W_i(\phi, \phi_i) = W\left(\frac{|\phi - \phi_i|}{\sigma}\right) - W\left(\frac{|\phi + \phi_i|}{\sigma}\right)\,.
\end{align}
The shape parameter $\sigma$ is tuned to ensure a good overlap of the basis functions. For the computations of this paper, we chose $\sigma \approx 0.41$. As discussed in the main text, we chose $M=11$ centres $\phi_i$ that are distributed in the region where the approximate kernel deviates most from the true kernel. To determine the coefficients $w_i$, we gathered $N=501$ linearly spaced collocation points in the same region and solved the resultant overdetermined linear system as in \labelcref{eq:LinearSystem} using a QR decomposition.

Finally, we note the present implementation is straightforward but not a unique choice and other methods may offer a significantly better performance. The optimisation of these auxiliary methods is, however, left for future work.

\section{PIKs as finite time corrections to the Langevin algorithm}
\label{sec:PIKFiniteTimeLangevinCorrection}

In \Cref{sec:PIKs}, we have seen how PIKs are constructed and how they can be used as generative models. By noting that generative models at their core facilitate a very fast thermalisation time for sampling algorithms, we discuss how PIKs conceptually take the role of finite time corrections to the Langevin algorithm which thermalises only in the infinite time limit. For related discussions, see also~\cite{Albergo:2025nets}. This connects the deterministic and continuous PIK model with the stochastic differential equation of the Langevin algorithm and is indeed analogous to the connection between continuous normalising flows and diffusion models~\cite{Song:2021sbgm}. 

For the Langevin algorithm, one considers the stochastic process
\begin{align}
    \frac{d\phi_t}{dt} = f_t(\phi) + g_t \; \eta_t\,,
\end{align}
where one uses the drift $f_t(\phi)$, the uncorrelated Gaussian noise $\eta_t$ with the pre-factor $g_t =\sqrt{2}$. This transformation induces the change in action at each time step according to
\begin{align}\label{eq:LangevinFokkerPlanck}
    \frac{d S_t(\phi)}{dt} = \frac{\partial}{\partial \phi} \, f_t(\phi) &- f_t(\phi) \cdot \frac{\partial}{\partial \phi} S_t(\phi) \nonumber \\[1ex]
    &+ \frac{\partial^2}{\partial \phi^2} \, S_t(\phi) - \left[\,\frac{\partial}{\partial \phi} S_t(\phi) \, \right]^2\;. 
\end{align}
As is known, in the infinite time limit, $t\to \infty$, one arrives at the stationary distribution $\propto e^{-S(\phi)}$ by choosing $f_t(\phi) = -\frac{\partial}{\partial \phi}  S(\phi)$. However, to arrive at the same distribution at some fixed and finite time, one can introduce a finite time correction to this approach via
\begin{align}
    f_t(\phi) = - \frac{\partial}{\partial \phi} S_t(\phi) + \dot{\phi}_t(\phi)\;,
\end{align}
where $\partial S_t(\phi)/\partial \phi$ is the drift term for the time-dependent action. Using this Ansatz in \labelcref{eq:LangevinFokkerPlanck}, one directly retrieves the Wegner equation as used in the main body of the text
\begin{align}
    \frac{d S_t(\phi)}{dt} = \frac{\partial}{\partial \phi} \, \dot{\phi}_t(\phi) - \dot{\phi}_t(\phi) \, \frac{\partial}{\partial \phi} S_t(\phi)\,.
\end{align}
From this point of view, one can see the PIKs as a constructive approach to accelerate the thermalisation time of the Langevin algorithm by introducing a finite time correction to the drift term.

\small
\bibliographystyle{apsrev4-2}
\bibliography{refs}

\end{document}